\documentclass[twocolumn,showpacs,preprintnumbers,amsmath,amssymb,prb,floatfix]{revtex4}

\usepackage{graphicx}
\usepackage{dcolumn}
\usepackage{bm}
\usepackage{epstopdf} 
 
\begin{document}

\title{Polar catastrophe, electron leakage, and magnetic ordering at the LaMnO$_3$/SrMnO$_3$ interface
}

\author{B. R. K. Nanda and S. Satpathy}
\affiliation{Department of Physics, University of Missouri, Columbia, Missouri 65211, USA
}

\date{\today}

\begin{abstract}

Electronic reconstruction at the polar interface LaMnO$_3$/SrMnO$_3$ (LMO/SMO) (100) resulting from the
polar catastrophe is studied from a model Hamiltonian that includes the double and super exchange interactions, the Madelung potential, and the Jahn-Teller coupling terms relevant for the manganites.  We show that
the polar catastrophe, originating from the alternately charged LMO layers and neutral SMO layers,
is quenched by the accumulation of an extra half electron per cell in the
interface region as in the case of the LaAlO$_3$/SrTiO$_3$ interface. In
addition, the Mn e$_g$ electrons leak out from the LMO side to
the SMO side, the extent of the leakage being controlled by the
interfacial potential barrier and the substrate induced epitaxial
strain.  The leaked electrons mediate a Zener
double exchange, making the layers adjacent to the interface ferromagnetic,
while the two bulk materials away from the interface retain their original type A or G
antiferromagnetic structures. A
half-metallic conduction band results at the interface, sandwiched by the two insulating bulks.
We have also studied how the electron leakage and consequently the magnetic ordering are affected by the  substrate induced epitaxial strain.   Comparisons are made with the results of the density-functional calculations for the (LMO)$_6$/(SMO)$_4$ superlattice.    
\end{abstract}


\maketitle

\section{Introduction}
Polar interfaces have commanded considerable interest recently because extra electrons  may migrate to the interface to ``heal" the polar catastrophe and these interfacial electrons may exhibit unusual two-dimensional properties. A polar interface of current interest is LaAlO$_3$/SrTiO$_3$ (LAO / STO)\cite{ohtomo2,brinkman,thiel,huijben,cen,satpathy} , which shows a variety of phenomena such as the Kondo resistance minimum, superconductivity,  magnetism, and metallic or insulating behavior under varying circumstances.\cite{brinkman}

\begin{figure}
\includegraphics[width=8.cm]{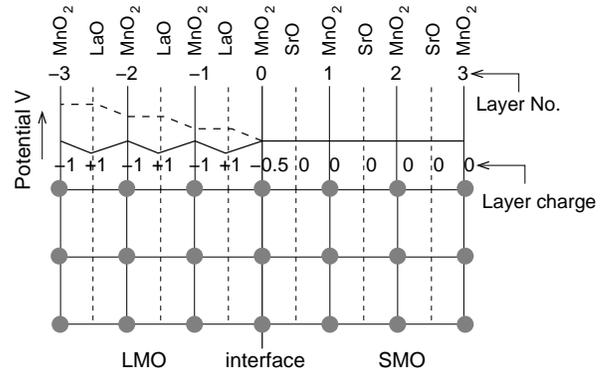}
\caption{Illustration of the polar catastrophe for the LMO/SMO (100) interface. The layers are neutral on the SMO side, but are charged $\pm 1$ on the LMO side leading to the unrestricted growth of the Coulomb potential (dashed line) and the scenario where it is healed  by accumulating 0.5 electrons per cell on the interfacial MnO$_2$ layer (solid line).  The layer numbers indicated in this figure are used throughout  the paper.}
\label{schema} 
\end{figure}

This paper is devoted to a theoretical study of the interface between LaMnO$_3$ and SrMnO$_3$, which is a polar interface (Fig. \ref {schema}) with the extra twist that both constituent materials are also magnetic. Presence of the Mn localized moments introduces a new interaction channel for the electron gas forming at the interface.
In fact, in a similar system, viz., a single $\delta$-doped LMO layer in a SMO matrix, where an extra electron per cell becomes introduced at the $\delta$-doped  layer, the formation of a spin-polarized two-dimensional electron gas was predicted due to the Zener double exchange between the itinerant electrons and the localized Mn moments.\cite{brk1} On the experimental front, it has been demonstrated that high-quality superlattices  of the manganites can be grown by MBE and the magnetic properties can be controlled by the superlattice period as well as by substrate-induced strain.\cite{anand, santos, aruta, anand1, yamada, brk2, brk3}
Various experimental and theoretical studies have found diverse magnetic phases for these superlattices.\cite{anand, santos, aruta, anand1, yamada, brk2, brk3, robertson, satoh, may, koida, salva, dong, lin}  

Both LMO and SMO are magnetic with LMO being a type A antiferromagnetic (AFM) insulator with the Mn-$t_{2g}^3e_g^1$ configuration
and consisting of alternating positively charged LaO and negatively charged MnO layers along the (100) direction. On the other hand, SMO is a type G AFM insulator with Mn-$t_{2g}^3e_g^0$ configuration with neutral SrO and MnO$_2$ layers. As a result of the layer charge configuration, a polar catastrophe - the divergence of the Coulomb potential away from the interface - arises in this system (Fig. \ref {schema}) as in the case of the LAO/STO interface. We note that unlike the latter case, where different layer termination produces two different n and p type interfaces, for the LMO/SMO interface, the MnO$_2$ layers being common to both, we just have a single type of interface, which by simple electron counting, is expected to be n type, with extra electrons coming to the interface region as illustrated in Fig. \ref {schema}.
We study the issues of polar catastrophe, charge leakage, and magnetism at the interface by using a tight-binding model Hamiltonian that includes all the relevant interactions in the system, complementing the study with the {\it ab initio} density-functional calculations. 
 
\begin{figure}
\hspace{-2cm}
\includegraphics[width=6.cm]{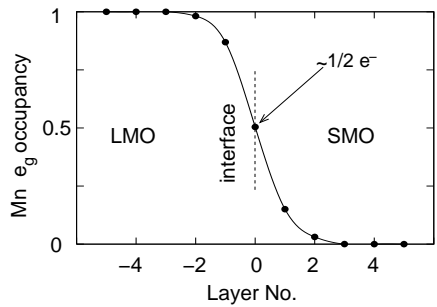}
\caption{\label{charge}Layer projected Mn e$_g$ occupancy across the  interface, as obtained from the model Hamiltonian with dielectric constant $\epsilon = 10$. About half an electron per Mn atom accumulate at the interface to remove the polar catastrophe and, furthermore, a small number of electrons leak from the LMO to the SMO side. Electrons at the interface form a spin-polarized 2DEG, sandwiched between the two insulating bulks.}
\end{figure}

The main results that emerge from our work are the following: (i)  
Half an electron per unit cell accumulates at the interface to avoid the polar catastrophe much like the case of LAO/STO, although now the electrons are much more confined to the interface (Fig. \ref{charge}). (ii) In addition, the Mn (e$_g$) electrons leak from the LMO to the SMO side, spreading to several layers. (iii) While the LMO and SMO layers away from the interface retain, respectively, the type A and type G  AFM  structures of the bulk, the interface becomes ferromagnetic (FM) due to the double exchange between the itinerant carriers and the Mn core spins.  (iv) Finally, we show that the magnetic ordering at the interface is sensitive to the substrate-induced epitaxial strain. However, our results suggest the absence of any canted magnetic state at the interface.

\section{Model Hamiltonian}
Density-functional calculations on the manganites have provided 
important insights into the electronic structure.\cite{satpathylmo,singh} The important electrons near the Fermi energy are the itinerant Mn (e$_g$) electrons, which interact with the localized Mn (t$_{2g}$) core spins via the double exchange.\cite{zener, andersen} The other important terms to consider are the  Jahn-Teller coupling of the e$_g$ electrons with the octahedral vibrational modes and the Madelung potential of the constituent atoms that would lead to the polar catastrophe without the electronic reconstruction at the interface. 

Incorporating these key interactions, we construct the model Hamiltonian 
\begin{eqnarray}
{\cal H} & = & \sum_{i\alpha}(\varepsilon_{i\alpha}+
\frac{1}{\epsilon}\sum_\nu M_{i\nu} q_\nu)
c_{i\alpha}^{\dagger}c_{i\alpha} \nonumber \\
& + & \sum_{\langle ij\rangle \alpha\beta}t_{i\alpha, j\beta}\cos(\theta_{ij}/2)(c^{\dagger}_{i\alpha}c_{j\beta} + H.c.) \nonumber \\
 & + & \frac{U}{2}\sum_i n_{i1}n_{i2} 
 + \sum_i {\cal H}_{JT}^i
+ \frac{J}{2}\sum_{\langle ij\rangle} \hat{S_i} \cdot \hat{S_j},
\label{htot}
\end{eqnarray}
which describes the motion of the Mn (e$_g$) electrons in a matrix of Mn core spins. These electrons are effectively spinless as discussed later. The e$_g$ electrons are restricted to the Mn sites, but  all atoms in the structure (Mn, Sr, La, and O) contribute to the electrostatic potentials that these electrons see, so that the Madelung matrix M$_{ij}$ goes over all atoms. 
 In the Hamiltonian above, $c^\dagger_{i\alpha}$  creates an electron at the $i^{th}$ Mn site and $\alpha$ is the orbital index of the e$_g$ electron ($x^2 - y^2$ or $z^2 - 1$), $n_{i\alpha}$ is the corresponding number operator, $\langle ij \rangle$ denotes summation over nearest neighbor pairs, and while the $i$ summation runs over only the Mn sites, the $\nu$ summation runs over all atoms in the structure. 
 
 The first term in the Hamiltonian describes the onsite energy, where the  energies $\varepsilon_{i\alpha}$ of the two e$_g$ orbitals could be split due to strain (considered later in the paper). The Madelung energy is given by 
 $E_{Mad} = \frac{1}{2 \epsilon}\sum_{\nu \nu^\prime}  M_{\nu \nu^\prime}  q_\nu q_{\nu^\prime}$, 
 where
  $q_\nu$ denotes the total charge (ionic + electronic) of the $\nu$-th atom, $\epsilon$ is the dielectric constant, the diagonal terms of the Madelung matrix $M_{\nu \nu^\prime}$ exclude the  Coulomb contribution from the same site, and the factor of two comes from double counting.  The Madelung potential seen by the e$_g$ electron is given by $V_{Mad} = d E_{Mad} / d n_{i\alpha} = \frac{1}{\epsilon}\sum_\nu M_{i\nu} q_\nu$,
  which appears in the first term in the Hamiltonian.
  The Madelung matrix for the structure is obtained by using the standard Ewald summation method.
 
 The second term in the Hamiltonian is the electronic hopping energy. We have taken the Hund's energy $J_H$ to be $\infty$, so that
the coupling between the core and the itinerant spins $ {\cal H}_{Hund} = - J_H \sum_{i\alpha}\vec{S_i} \cdot \vec{s}_{i\alpha}$ makes the electron state inaccessible, where the electron spin is anti-aligned with the local $t_{2g}$ core spin. Thus the itinerant electrons are effectively  spinless and the electron hopping is diminished by the Anderson-Hasegawa $ \cos (\theta_{ij}/2)$  factor\cite{andersen}, where $\theta_{ij}$ is the angle between the (classical) core spins at the two sites. The nearest-neighbor hopping integral $t$ depends on the relative positions of the two Mn sites in the lattice. For nearest-neighbor hopping in the xy-plane or along the z-axis, we have:
\begin{equation}
\begin{array}{cc}
t_{\alpha\beta}^{xy} = 
\frac{V_{\sigma}}{4}\left(\begin{array}{cc}    
1 & -\sqrt{3}\\
-\sqrt{3} & 3\\
\end{array} \right),&
t_{\alpha\beta}^{z} = 
V_{\sigma}\left(\begin{array}{cc}    
1 & 0\\
0 & 0\\
\end{array} \right),
\end{array} 
\end{equation}
where $V_{\sigma}$ is the dd$\sigma$ matrix element and we have neglected the much smaller dd$\delta$ interaction. The third term in the Hamiltonian is the Coulomb repulsion between the two e$_g$ orbitals on the same site.

The Jahn-Teller (JT) coupling term on each site is given by
\begin{equation}
{\cal H}_{JT}  =  \frac{1}{2}K(Q_2^2 + Q_3^2) -g(Q_3\tau_z + Q_2\tau_x),
\label{hjt}
\end{equation}
where $\vec{\tau}$ is the pseudospin describing the two e$_g$ orbitals, viz., $|\uparrow\rangle$ = $|z^2 - 1\rangle$ and $|\downarrow\rangle$ = $|x^2 - y^2\rangle$. With the corresponding creation operators being $c_{i1}^\dagger$ and 
$c_{i2}^\dagger$, respectively, we have:  $\vec{\tau} = \sum_{\alpha \beta} c_{i\alpha}^\dagger  \vec{\tau}_{\alpha \beta} c_{i\beta}$, where the greek indices denote the orbitals and $i$ is the site index.  
The quantities $Q_2$ and $Q_3$ are, respectively, the basal plane distortion mode and the octahedral stretching mode at the i-th site, $K$ is the elastic stiffness constant, and $g$ is the JT coupling strength. The final term in the Hamiltonian describes the AFM superexchange interaction between the Mn t$_{2g}$ core spins.

The typical values of the Hamiltonian parameters used in our calculations are: K = 9 eV/{\AA}$^2$ following from the optical studies on La$_{0.85}$Sr$_{0.15}$MnO$_3$\cite{millis1}, g = 2.0 eV/{\AA}, and V$_{\sigma}$ = -0.5 eV following earlier density-functional results\cite{popovic},
$J = 26$ meV, estimated from the Ne\'el temperature of CaMnO$_3$,\cite{wollan, rushbroke} a compound similar to SrTiO$_3$, and  finally, U = 3 eV, which is a reasonable for the Mn (e$_g$) electrons. The values for the JT distortions in the LMO bulk are:\cite{elemans} $Q_2^0 = 0.28$  {\AA} and  $Q_3^0 = -0.10$ {\AA}. These distortions are bulk like on the LMO side and go to zero on the SMO side, where there are no e$_g$ electrons. There is a transition region in between, where, for simplicity, we have taken the distortion  strengths to be linearly dependent on the site electron occupancy of the e$_g$ orbitals, which would be the result for the isolated octahedron.  This is easily seen from an examination of the single-site JT Hamiltonian Eq. \ref{hjt}.
If the electron occupancy $n$ is taken to be a continuous variable  as appropriate for a mean-field model, it immediately follows from the diagonalization of  the Hamiltonian (\ref{hjt}) that the optimized magnitude of the distortion is given by: $(Q_2^2+Q_3^2)^{1/2}=gn/K$. Following this argument, we use the distortion magnitudes at each Mn site as: $Q_{2} = n \times Q_2^0$, where $Q_2^0$ is the magnitude for bulk LMO and similarly for $Q_3$.  

There is one more point to be made regarding the model. The measured value of the dielectric constant for LMO is 16-18,
while for SMO, it is roughly 35.\cite{LMO-dielectric1,LMO-dielectric2} From work on the LAO/STO interface, it has been argued that the dielectric constant for STO is drastically reduced\cite{copie}  in the presence of an electric field, which exists at the interface.
We however use a uniform dielectric constant  in our model calculations following earlier authors,\cite{Siemons} which is reasonable within the spirit of our model.
We find that the electron density profile near the interface region obtained from our model using  $\epsilon \approx 8-10$ fits well with the {\it ab initio} density-functional results. Unless otherwise stated, we have taken $\epsilon  = 10$ in our model calculations.

We solve the model Hamiltonian within the Hartree-Fock mean field approximation.  A supercell geometry is adopted for the convenience of calculation; we find that the (LMO)$_{10}$/(SMO)$_{10}$ cell is large enough for our purposes.
For a fixed core spin configuration, beginning with an initial set of Mn site occupancies, the band structure was computed in the Brillouin zone, from which new site occupancies were determined, and the process was repeated until self-consistency was reached. 

To compare with the results of our model calculations, we have also performed an {\it ab initio} density functional calculation using the linear muffin-tin orbitals method (LMTO) with the generalized gradient approximation\cite{gga} to the exchange-correlation functional with the Coulomb correction (GGA + U) using a somewhat smaller supercell, viz., (LMO)$_6$/(SMO)$_4$, and a value of U = 3 eV, which is reasonable for the Mn (e$_g$) electrons.

\section{Results and Discussions}

\begin{figure}
\includegraphics[width=7cm]{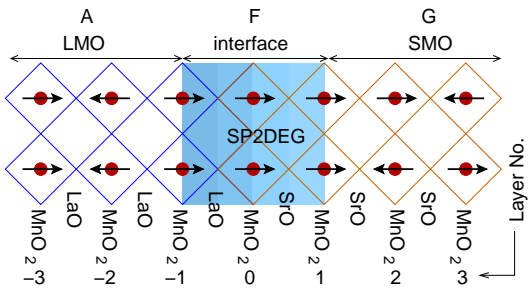}
\caption{\label{magfig} (Color online) Minimum-energy magnetic configuration at the LMO/SMO interface for typical parameters and unstrained condition as obtained from both the density-functional and model calculations. The first MnO$_2$  layer on the SMO side can be tuned FM or AFM by changing the strain condition as discussed in the text.}
\end{figure}

We examined the computed total energy for different configurations of the core Mn spins and found that the minimum energy structure is one where away from the interface, the G and A type antiferromagnetism of the bulks are recovered, while the interface region becomes ferromagnetic. 
The magnetic configurations at the interface are characterized by magnetic ordering within each plane, indicated by unbracketed symbols in Table I, and between the adjacent planes, which are indicated by bracketed symbols. Deeper in the bulks, both LMO and SMO are AFM type A and G, respectively.
The computed energies for different magnetic configurations are listed in Table I and the configuration with the lowest energy is illustrated in Fig. \ref{magfig}. The results presented in the following sections correspond to the lowest-energy configuration, unless otherwise stated, and we note that quantities such as charge reconstruction and Madelung potentials are rather insensitive to the magnetic configuration.

\begin{table}
\caption{Relative energies (per Mn interface atom) for different magnetic configurations. The intra-layer magnetic couplings are denoted by the symbols F or A (FM or AFM ordering within the plane), while bracketed symbols (F) or (A) denote the coupling between the adjacent interface layers. (F, A) means that half the interlayer bonds are FM and the other half are AFM. The lowest energy structure, corresponding to the first line of the Table, is shown in Fig. \ref{magfig}.
}
\begin{center}
\begin{tabular}{ccccc|c|c}
\hline
\multicolumn{4}{c}{Magnetic Order at the Interface}&&\multicolumn{2}{c}{Energy}\\
\hline
intra-&inter-&intra-&inter-&intra-&DFT&Model\\
layer&layer&layer&layer&layer&&\\
Layer$\#$-1&&Layer$\#0$&&Layer$\#$1 &(eV)&(eV)\\ 
\hline

F&(F)&F&(F)&F&0&0\\
F&(F)&F&(F, A)&A&0.162&0.406\\
F&(A)&F&(F)&F&0.101&0.175\\
F&(A)&F&(F, A)&A&0.295&0.684\\
F&(F, A)&A&(A)&A&0.957&1.026\\
\hline
\end{tabular}
\end{center}
\end{table}
\subsection{Polar catastrophe and Interfacial charge reconstruction} 

\begin{figure}
\includegraphics[width=6cm]{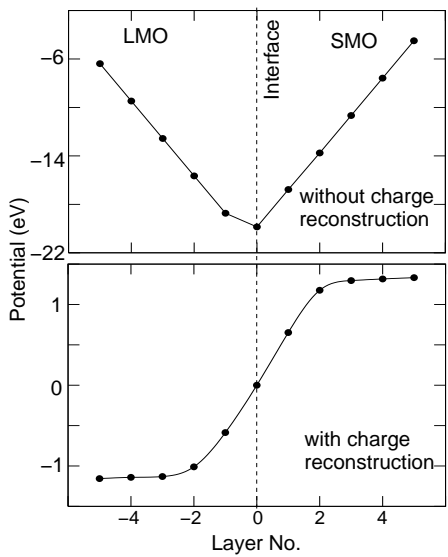}
\caption{\label{madelung} Madelung potential seen by the Mn ions in various layers without and with the charge reconstruction as obtained from the self-consistent calculation using the model Hamiltonian (Eq. 1). Without the charge reconstruction, the potential grows unrestricted away from the interface (polar catastrophe), which is healed by accumulation of half an electron per cell at the interface. The reconstructed charge at the interface has a dipole moment that leads to a potential step (bottom figure).}
\end{figure}

 As widely discussed in the context of the LAO/STO interface, alternately stacked positive and negative layers - in the present case LaO and MnO$_2$ layers respectively -   
lead to a divergent Coulomb potential, the so called polar catastrophe. In Fig. \ref{madelung}, we have plotted the initial Madelung potential of the charged layers (taking the nominal charged states for all atoms) and the final potential after charge reconstruction has occurred and self-consistency has been achieved.

\begin{figure}
\includegraphics[width=7cm]{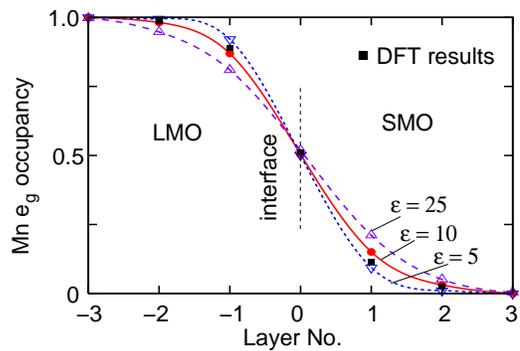}
\caption{\label{valence}   (Color online) 
Mn e$_g$ occupancy across the LMO/SMO interface for different values of the dielectric constant $\epsilon$  (symbols connected by lines) compared with the density-functional results  for the (LMO)$_6$/(SMO)$_4$ superlattice (filled squares). An increased $\epsilon$ results in a diminished Madelung potential and, consequently, electrons leak deeper into the bulk.
}
\end{figure}

	Charge reconstruction at the interface occurs in two ways as may be seen from Fig. \ref{valence}. First, a monopole charge, half electron per cell, or $3.4 \times 10^{14} \ e$/cm$^2$ accumulates at the interface region to cancel the original electric field discontinuity at the interface. A significant portion of these electrons reside on the interfacial MnO$_2$ layer as seen from the figure. The interface region demands this monopole charge in order to quench the polar catastrophe. In our supercell geometry, which has two identical interfaces per supercell and also an extra electron per cell after satisfying the nominal ionic charges of all atoms, this electron was quite conveniently shared between the two interfaces.  Thus we have half an electron per cell available at each interface, exactly the amount needed to quench the polar catastrophe. In the experimental situation, this monopole charge must come from the surfaces and/or bulk defect states as necessary, so that the demand for the monopole charge at the interface can be met.
	  A second aspect of the charge reconstruction is that the Mn (e$_g$) electrons leak from the LMO side  into the SMO side, creating a dipole moment at the interface, which results in a potential discontinuity of about 2 eV as seen from the bottom part of Fig. \ref{madelung}.
	The magnitude of this surface dipole moment density is estimated to be  $p \approx 0.11 e  /{\rm \AA}$.

\begin{figure}
\includegraphics[width=5cm]{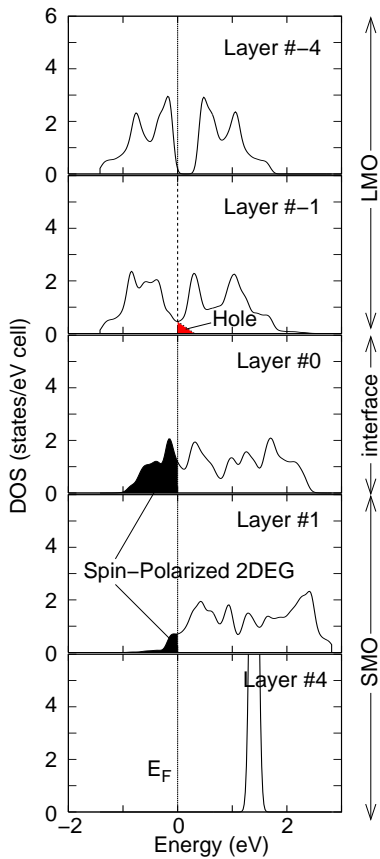}
\caption{\label{dos}  (Color online)  Layer projected Mn e$_g$ densities of states  obtained from the model Hamiltonian indicating extra electrons or holes accumulated at the interfacial layers, while away from the interface, the e$_g$ states are either half filled (LMO) or unoccupied (SMO). The electrons are effectively spinless, with only the spin state aligned parallel to the Mn core spin at a particular Mn site allowed to be occupied ($J_H=\infty$), so that we have a single spin channel in this figure. }
\end{figure}

The interfacial electronic reconstruction is consistent with the layer projected densities of states (DOS) shown in Fig. \ref{dos}. Here we see that at the interfacial MnO$_2$ layer, approximately one fourth of the e$_g$ states are occupied indicating a net occupancy of about half an electron per Mn atom. The DOS for the layers away from the interface resemble the bulk electronic configuration of the respective materials. Layers close to the interface have either a small number of electrons in the e$_g$ bands (SMO side) or a small number of electrons missing (LMO side), consistent with the  electron distribution of Fig. \ref{valence}. 

\subsection{Density-functional results and half-metallic behavior }
\begin{figure}
\includegraphics[width=6.5cm]{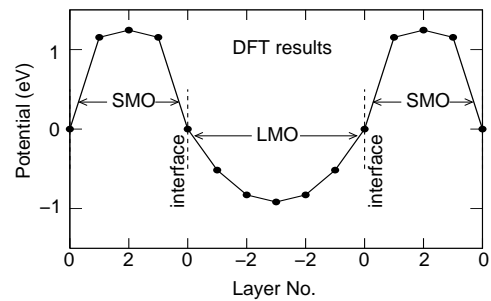}
\caption{\label{dftpot} Variation of the energy of the lowest Mn e$_g$ state of each MnO$_2$ layer, obtained using DFT-LMTO from the layer-projected wave-function characters for the (LMO)$_6$/(SMO)$_4$ superlattice. This variation in energy is indicative of the potential experienced by the Mn e$_g$ electrons across the interface.}     
\end{figure}

In this subsection, we report the results of our density-functional theory (DFT) calculations, which validate the results of our model and also address the issue of half metallicity of the conduction bands at the interface. 
Half-metallic behavior, where one spin band is metallic, while the other is insulating, was  already predicted for the delta doped LMO/SMO structure, where a single LMO layer is doped in a matrix of SMO.\cite{brk1} There, the extra electrons coming from the LMO layer became confined in the electrostatic potential well of the LMO layer producing a ferromagnetic alignment of the core Mn spins at the interface and in turn became spin polarized owing to the Zeeman field of the Mn moments. Our DFT studies show that we have a similar situation for the LMO/SMO interface,  with the difference that  now the interfacial electrons originate from the polar catastrophe rather than from the La dopants in the delta doped structure.

The DFT calculations were performed for the (LMO)$_6$/(SMO)$_4$ superlattice  using the linear muffin-tin orbital (LMTO) with the  gradient approximation for the exchange correlation functional and including the on-site Coulomb term (GGA+U)  with U = 3 eV.
Both the in-plane and out-of-plane lattice parameters were taken to be the average lattice parameters of LMO and SMO. Lattice relaxation effects were not included, which are expected to cause the electrons to spread somewhat further into the bulk away from the interface, but otherwise not change the essential physics of the problem. Effect of lattice relaxation for a similar system, viz.,  delta-doped SrTiO$_3$/(LaTiO$_3)_1$/SrTiO$_3$ was discussed in our earlier work using detail density functional calculations.\cite{Larson}

In Fig. \ref{dftpot}, we have shown the relative potential seen by the Mn e$_g$ electrons at each MnO$_2$ layer as obtained from the DFT calculations. The variation in the potential across the interface is obtained by calculating the energy of the lowest Mn e$_g$ band state in each MnO$_2$ layer, which can be obtained from the layer-projected wave-function characters. From the figure, we see that the magnitude of the potential discontinuity at the interface (about 2 eV) is similar to the one obtained from the model calculation (Fig. \ref{madelung}). 

\begin{figure}
\includegraphics[width=8.5cm]{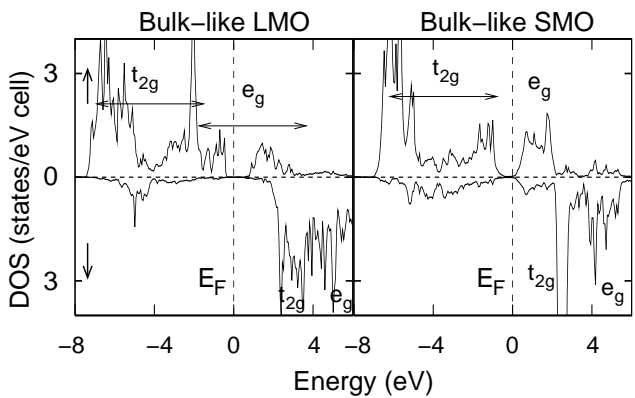}
\caption{\label{dos1} Density-functional results for the majority and minority spin Mn-d densities of states for the SMO and LMO layers away from the interface indicating the bulk-like insulating behavior. Results are for the innermost MnO$_2$ layers in the (LMO)$_6$/(SMO)$_4$ superlattice.}
\end{figure}
%

\begin{figure}
\includegraphics[width=6cm]{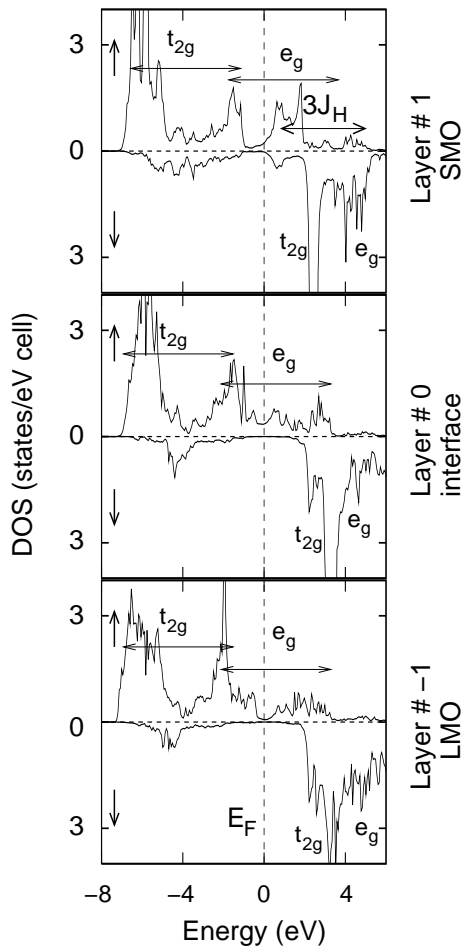}
\caption{\label{dos2} Same as Fig. \ref{dos1} for interfacial layers, indicating the presence of carriers at the interface, which are also spin-polarized. Spin-minority states are unoccupied indicating half-metallicity.
}
\end{figure}

The layer projected densities of states for the LMO)$_6$/(SMO)$_4$ superlattice, as obtained from the DFT calculations, are shown in Figs. \ref{dos1} and \ref{dos2}. From the figures, we see that the DOS for the innermost MnO$_2$ layers show bulk-like behavior, while the occupation of the   Mn e$_g$ states for the interfacial layers are very similar to the results obtained from our model calculation (Fig. \ref{dos}).

A notable feature of the density functional results is that, for each layer the minority-spin states are completely unoccupied at the Fermi level, leading to the  half-metallic behavior. In other words the two dimensional electron gas formed at the interface is completely spin-polarized. We note that this behavior is obviously absent in the widely studied LAO/STO interface, which lacks any magnetic atoms.

\subsection{Effect of strain on electron leakage and magnetic ordering}

\begin{figure}
\includegraphics[width=8.5cm]{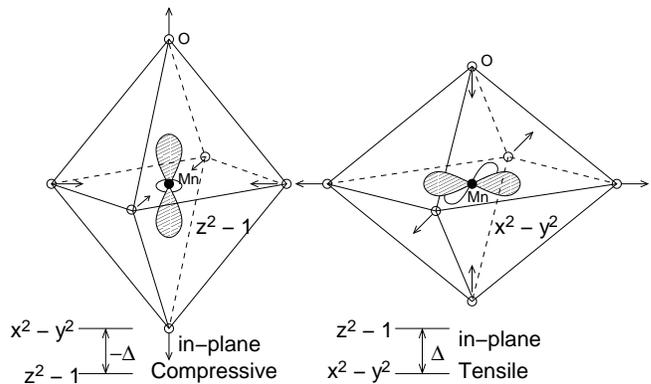}
\caption{\label{strain} Energy splitting of the Mn  e$_g$ orbitals due to in-plane compressive and tensile strain conditions used in our model Hamiltonian. }
\end{figure}

In this subsection, we study the effect of strain on the electron leakage across the 
interface, which in turn affects the magnetism. Density-functional calculations have shown that strain alters the relative energy between the two e$_g$ orbitals, causing a change in the orbital ordering.\cite{brk3} A change of the symmetry of the occupied state from $x^2-y^2$ to $z^2-1$ would increase electron hopping across the interface leading to an increase of electron leakage. Strain is taken into account in our model Hamiltonian (\ref{htot}) via the on-site energy:
\begin{equation}
\varepsilon_{i\alpha} = \left 
\lbrace \begin{array}{ll}0 \ \ \ \ (\alpha = x^2-y^2) \\
\Delta \ \ \ \ (\alpha = z^2-1) \end{array} 
\right.,
\end{equation}
where a positive $\Delta$ correspond to an in-plane tensile strain (see Fig. \ref{strain}).

 
%

\begin{figure}
\includegraphics[width=7cm]{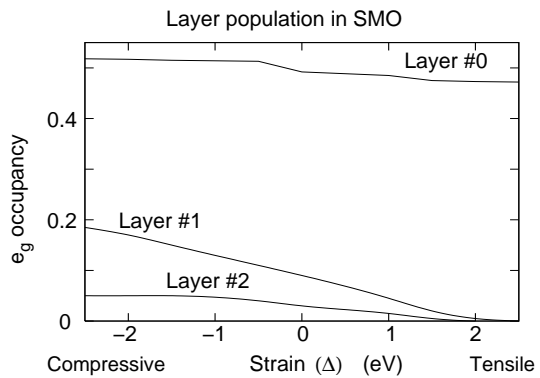}
\caption{\label{occupancy} Electron leakage to the individual MnO$_2$ layers on the SMO side as a function of  strain. }
\end{figure}
 The computed layer occupancy of the electrons is plotted in Fig. \ref{occupancy}, where we have shown the electron leakage to various Mn layers in the SMO side. As expected, the layer occupancy diminishes with increasing $\Delta$, a result of the lower on-site energy for the $x^2-y^2$ orbital, which leads to a larger $x^2-y^2$ character of the occupied e$_g$ electrons and consequently to diminished electronic hopping across the interface. The layer occupancy of the interface layer \# 0 remains close to 0.5 e$^-$, the density needed to quench the polar catastrophe. 
\begin{figure}
\includegraphics[width=7cm]{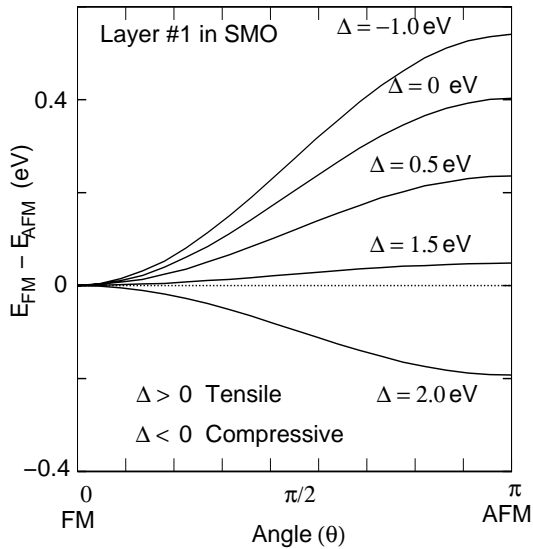} 
\caption{\label{canting} The total energy as a function of the canting angle $\theta$ and strain $\Delta$. The parameter $\theta$ is the angle between the neighboring Mn core spins in the first MnO$_2$ layer on the SMO side. The figure shows that, the considered MnO$_2$ plane either stabilizes with FM order or AFM order depending on the strain condition. No canted magnetism is found.}
\end{figure}
%
\begin{figure} [b]
\includegraphics[width=7cm]{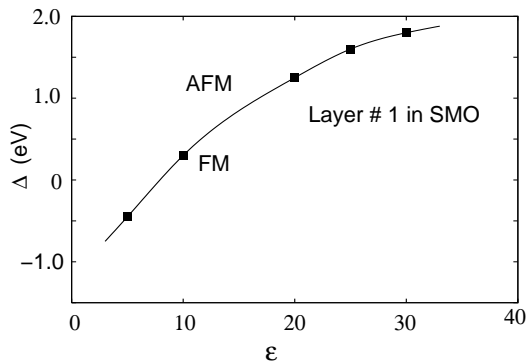} 
\caption{\label{magneticphase} Magnetic phase diagram for the first MnO$_2$ layer on the SMO side as a function of the dielectric constant $\epsilon$ and  strain $\Delta$. Rest of the layers maintain the magnetic behavior of the corresponding bulk materials, either AFM or FM within each layer, throughout the range of parameters considered. }
\end{figure}

The magnetism in the manganites is determined by a competition between the superexchange between the core t$_{2g}$ spins and the double exchange between the core spins and the itinerant e$_g$ carriers. The strength of the latter clearly depends on the concentration of the itinerant carriers. In SMO bulk, the AFM super exchange is the only term, since there are no e$_g$ electrons, which leads to a  N\'{e}el type G order. As the concentration of the e$_g$ electrons is increased, the competition between the super and the double exchange could lead to a canted magnetic state, eventually resulting in a ferromagnetic state if the double exchange dominates.\cite{andersen}

To study if strain can affect the magnetic behavior by modifying the electron leakage, we have solved the Hamiltonian with different orientation of the Mn core spins in order to obtain the ground-state magnetic configuration. Canted states were also considered in addition to the FM and AFM states. Within the range of strain studied, the strain being parametrized by the energy splitting parameter $\Delta$, we find that the magnetic ground state does not change from the one shown in Fig. \ref{magfig}, except for the magnetic configuration of the layer \# 1, which can be altered between FM and AFM depending on the strain condition. In particular, the electron leakage to this layer can be large enough to produce a net ferromagnetic interaction.

Fig. \ref{canting} shows the total energy as a function of the canting angle $\theta$ between neighboring core spins in the MnO$_2$ layer \# 1. There is a transition between the FM and AFM states, but we do not find the occurrence of a canted state within our model.
Since the electron leakage beyond this layer is small, we see the bulk magnetic ordering beyond the first layer. Fig. \ref{magneticphase}  shows the magnetic phase diagram for the same MnO$_2$ layer  as a function of strain and the dielectric constant.

\section{Summary}

In summary, by solving a model Hamiltonian in the Hartree-Fock mean-field theory, we studied the charge reconstruction at the polar interface of LMO/SMO. The results were complemented by {\it ab initio} density-functional calculations of the (LMO)$_6$/(SMO)$_4$ superlattice.
Two types of electronic reconstructions were found at the interface. First, there is an accumulation of an extra half electron per cell in the
interface region as in the case of the LAO/STO interface in order to quench the 
polar catastrophe. Secondly, the Mn e$_g$ electrons leak out from the LMO side to
the SMO side and alter the magnetism at the interface, while away from the interface, the magnetism of the respective bulk materials is preserved. Our calculations suggest the presence of a half-metallic two-dimensional electron gas in the interfacial region under certain strain conditions. Indeed, experimental evidence for half-metallicity for this interface was recently obtained from magneto-optical Kerr effect.\cite{Ogawa} We note that the LMO/SMO interface is a magnetic counterpart of the much studied polar interface between LAO and STO and as such presents an extra degree of freedom for the study of the two-dimensional electron physics.

This work was supported by the U. S. Department of Energy through Grant No.
DE-FG02-00ER45818.


\end{document}